\documentclass[conference,a4paper]{IEEEtran}
\IEEEoverridecommandlockouts
\usepackage{cite}
\usepackage{amsmath,amssymb,amsfonts}
\usepackage{algorithmic,algorithm}
\usepackage{graphicx}
\usepackage{textcomp}
\usepackage{xcolor}
\def\BibTeX{{\rm B\kern-.05em{\sc i\kern-.025em b}\kern-.08em
    T\kern-.1667em\lower.7ex\hbox{E}\kern-.125emX}}

\usepackage{pgf}
\usepackage{tikz}
\usepackage{pgfplots}
\pgfplotsset{compat=1.18}
\usetikzlibrary{shapes.misc,shapes,external,pgfplots.statistics,automata,arrows,backgrounds,positioning}

\usepackage{xspace}
\usepackage{siunitx}
\usepackage{url}

\usepackage[hidelinks]{hyperref}
\usepackage{orcidlink}

\begin{document}

\title{Core QUIC: Enabling Dynamic, Implementation-Agnostic Protocol Extensions %
\thanks{This paper is an extended, technical report of the publication at the 2024 IFIP Networking's SLICES workshop. If you consider citing this work, please refer to the SLICES workshop version.}
}

\author{\IEEEauthorblockN{Quentin De Coninck \orcidlink{0000-0001-6483-3157}}
\IEEEauthorblockA{\textit{University of Mons (UMONS)} \\
Mons, Belgium \\
quentin.deconinck@umons.ac.be}
}

\maketitle

\newcommand{\protocoloperation}{protocol routine\xspace}
\newcommand{\protocoloperations}{protocol routines\xspace}

\newcommand{\pluginop}{\texttt{pluginop}\xspace} %
\newcommand{\pluginopcommon}{\texttt{pluginop-common}\xspace} %
\newcommand{\pluginopmacro}{\texttt{pluginop-macro}\xspace} %
\newcommand{\pluginopwasm}{\texttt{pluginop-wasm}\xspace} %
\newcommand{\pluginoperation}[1]{\textit{\textbf{#1}}}
\newcommand{\cquic}{Core QUIC\xspace}
\newcommand{\pquic}{PQUIC\xspace}
\newcommand{\xbgp}{xBGP\xspace}
\newcommand{\wasm}{Wasm\xspace}

\newcommand{\anchor}[1]{\textbf{\textsc{#1}}}

\begin{abstract}
  While applications quickly evolve, Internet protocols do not follow the same pace.
  There are two root causes for this.
  First, extending protocol with cleartext control plane is usually hindered by various network devices such as middleboxes.
  Second, such extensions usually require support from all participating entities, but often these run different implementations, leading to the chicken-and-egg deployment issue.
  The recently standardized QUIC protocol paved the way for dealing with the first concern by embedding encryption by design.
  However, it attracted so much interest that there is now a large heterogeneity in QUIC implementations, hence amplifying the second problem.

  To get rid of these deployment issues and to enable inter-operable, implementation-independent innovation at transport layer, we propose a paradigm shift called Core QUIC.
  While Core QUIC keeps compliant with the standardized QUIC protocol, it enforces implementation architecture such that any Core QUIC-supporting participant can be extended with the same, generic bytecode.
  To achieve this, Core QUIC defines a standardized representation format of common QUIC structures on which plugins running in a controlled environment can operate to extend the underlying host implementation.
  We demonstrate the feasibility of our approach by making two implementations Core QUIC-compliant.
  Then, we show that we can extend both with the same plugin code over several use cases.
\end{abstract}

\section{Introduction}

  The story of the Internet showed that deploying new features in end-to-end protocols over the Internet is often difficult.
  Two main elements affect this.
  First, there are many middleboxes in the network affecting the behavior of communicating protocols.
  In-network devices often interact with the traffic based on its public metadata, such as packet headers.
  Extending TCP was shown to be difficult~\cite{honda_is_2011} since negotiated options are observable by in-network third-parties such as firewalls that may block them.
  Legacy network nodes also prevented protocols such as SCTP~\cite{budzisz_taxonomy_2012} from being broadly deployed.
  
  Second, there is strong heterogeneity in the participating devices, coming from different vendors and shipping with diverse hardware and software.
  The Internet built its interoperability success thanks to open standards defining protocol messages and actions to take.
  Still, it enables very different implementation designs and transport protocols are often implemented in the operating system.
  Defining protocol extensions %
  therefore requires willingness, cooperation and implementation efforts from the various implementations' maintainers, a process that usually takes years, when it succeeds~\cite{wirtgen_xbgp_2023}.

  Recently, the network community tacked the first middlebox issue by introducing QUIC~\cite{rfc9000}, a UDP-based transport protocol with built-in encryption of not only the carried data, but also its protocol metadata.
  Thanks to this encryption, all control information, including extension negotiation, cannot be altered by middleboxes, making QUIC resilient to their interference.
  QUIC attracted a lot of interest, and there is now more than twenty publicly known implementations. %
  This success reintroduces the previously discussed heterogeneity issue.
  To reach wide deployment, most of these implementations need to agree on the proposed feature, and then integrate it and deploy to devices.
  Only a few large actors can sustain such effort, constraining innovation into their hands.

  To enable inter-operable, implementation-independent innovation at transport layer, we propose a paradigm shift called \emph{\cquic}.
  While \cquic keeps compliant with the QUIC standard~\cite{rfc9000}, it enforces implementation architecture such that \textbf{any \cquic-supporting participant can be extended with the same plugin} consisting in architecture-independent bytecode.
  In addition, \cquic addresses deployment and implementation concerns by enabling partial support and does not require major changes in the implementation's internals.
  To achieve this, Core QUIC defines a standardized representation format of common QUIC structures on which plugins running in a controlled environment can operate to extend the underlying host implementation.
  By making \textbf{two different implementations \cquic-compliant}, we demonstrate the feasibility of our approach.

The paper is organized as follow.
First, Section~\ref{sec:background} introduces background on QUIC.
Then, Section~\ref{sec:design} details the architecture of \cquic.
We describe our \cquic library implementation in Section~\ref{sec:implementation} and benchmark it in Section~\ref{sec:microbenchmarks}.
We demonstrate how plugins can extend our two \cquic implementations with several use cases in Section~\ref{sec:usecase}.
Finally, Section~\ref{sec:related} discusses related works and Section~\ref{sec:conclusion} concludes the paper.

\section{Background on QUIC}\label{sec:background}

Our work focuses on QUIC, a transport protocol providing reliable and encrypted services atop UDP.
To setup the encryption, QUIC connections rely on TLS 1.3~\cite{rfc8446}.
During the TLS handshake, QUIC endpoints advertise \emph{transport parameters} to communicate flow-control values as well as their support for specific QUIC features and extensions. 
Unlike TCP, except for a connection identifier in the header that remains in clear-text for routing purposes, the whole QUIC packet (i.e., most of the header and the payload) is encrypted and authenticated.
The encrypted payload contains \emph{frames}.
These form the root QUIC messages, and they follow a type-value format.
While the core specification allocates 30 values~\cite{rfc9000}, there are $2^{62}$ possible type values.
While the STREAM frames carry the application data, most of the frames are QUIC control information.
The ACK frame acknowledges to the peer the reception of QUIC packets.
The MAX\_DATA frame advertises the flow-control limits.
An endpoint can probe a network path using the PATH\_CHALLENGE frame, expecting its peer to send back a PATH\_RESPONSE.
The PADDING frame increases the packet's size without containing any information and is usually used to perform path MTU discovery.

\section{Designing Core QUIC}\label{sec:design}

The design of Core QUIC focuses on the easiness to pluginize the implementation.
Specifically, the effort required by the QUIC maintainer to include all the pluginization mechanisms should be as minimal as possible.
The easier the integration is, the higher the potential adoption of the system.
Specifically, \cquic has the following design objectives.

\textbf{\cquic can dynamically load features on a per-connection basis, regardless of the QUIC host implementation.}
The fundamental idea behind \cquic is to have implementations that can be tuned or extended without requiring binary change.
While PQUIC~\cite{de_coninck_pluginizing_2019} shares a similar idea, the development of its plugins is strongly tight to the PQUIC implementation. %
Instead, \cquic provides a common layer where any compliant implementation can be safely extended through the same plugin.
Such a layer defines \emph{routines} and \emph{fields} that any QUIC implementation needs to expose.

\textbf{Implementations can incrementally support \cquic.}
The story of the Internet taught us that successful solutions need to be easily deployable.
Given the paradigm shift of \cquic, such a solution should be simple to integrate.
Furthermore, QUIC implementations have very different internal architectures and some QUIC features may be harder to expose.
To tackle this, initial \cquic support requires minimal changes to the QUIC implementation.
The common layer exposition can then be incrementally implemented, and \cquic checks at plugin load time that all the requested elements
are actually provided by the host implementation. %

\textbf{\cquic plugins operate in a safe architecture-independent environment.}
Plugins consist in bytecode that may have been written by anyone.
They may contain unintentional mistakes or malicious content.
To mitigate this, plugins operate in an isolated environment and cannot access memory outside of its scope.

\textbf{\cquic supports the combination of non-overlapping plugins.}
In both TCP and QUIC, many extensions can be simultaneously enabled on a given session.
Indeed, these usually provide orthogonal features and, when applicable, define extension-specific messages.
\cquic keeps this property at the plugin level, as long as plugins do not override the exact same QUIC routine.
Furthermore, these plugins can collaborate through a well-defined interface.

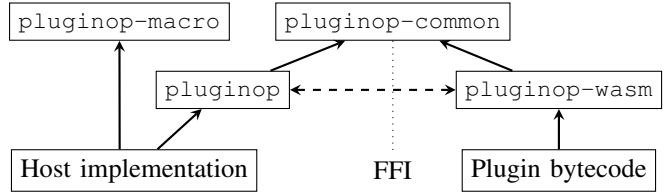
\begin{figure}
	\centering
	\begin{tikzpicture}
		\tikzstyle{arrow} = [thick,->,>=stealth]
		
		\node[rectangle, draw] (common) {\small \pluginopcommon};
		\node[rectangle, draw, left=1.7em of common] (macro) {\small \pluginopmacro};

        \node[below=1.5em of common] (anchor){};
        \node[below=2em of anchor] (ffi){FFI};
		
		\node[rectangle, draw, left=3.5em of anchor] (core) {\small \pluginop};
		\node[rectangle, draw, right=2em of anchor] (wasm) {\small \pluginopwasm};
		
		\node[rectangle, draw, left=4em of ffi] (impl) {Host implementation};
		\node[rectangle, draw, right=1.5em of ffi] (plugin) {Plugin bytecode};
		
		\draw[arrow] (impl.124) -- (macro);
		\draw[arrow] (impl) -- (core);
		\draw[arrow] (core) -- (common);
		\draw[arrow] (plugin) -- (wasm);
		\draw[arrow] (wasm) -- (common);
        \draw[dotted] (common) -- (ffi);
        \draw[arrow, dashed, <->] (core) -- (wasm);
		
	\end{tikzpicture}
	\caption{Architectural view of the different components of \pluginop. Plain arrows mean "uses", dashed arrows mean "interacts".}
	\label{fig:overview-architecture}
\end{figure}

To achieve this, this paper introduces \pluginop, a library enabling QUIC implementations to be dynamically extended through plugins.
Figure~\ref{fig:overview-architecture} shows the four architectural components. %
The core \pluginop library is the one a QUIC implementation needs to integrate to become \cquic-compliant.
The remaining of this Section describes how these building blocks work together to define the \protocoloperations (Section~\ref{sec:protocol-operations}), to set up a common representation layer (Section~\ref{sec:common-representation}), to run plugins in a safe environment (Section~\ref{sec:running-plugins}) and to integrate \cquic in an existing QUIC implementation (Section~\ref{sec:making-cquic-compliant}).

\subsection{Defining Protocol Routines} \label{sec:protocol-operations}
To provide programmability, \cquic defines hooks inside QUIC implementations.
Such hooks are called \emph{\protocoloperations} and correspond to operations that any QUIC implementation must provide.
Each of these \protocoloperations define input parameters and, optionally, output values.
The definition of such \protocoloperations introduces a trade-off.
On the one hand, having numerous \protocoloperations enables fine-grained protocol tuning and extensibility.
On the other hand, requiring support for a large number of \protocoloperations makes it harder for implementations to fully support \cquic, especially when dealing with different software architectures.
\cquic \protocoloperations should be common to all implementations.

As \cquic's initial goal focuses on providing extensions to any compliant QUIC implementation with a single bytecode, we put our attention on how QUIC negotiates them.
QUIC endpoints advertise parameters and support for extensions through QUIC transport parameters, exchanged during the handshake.
Such transport parameters follow a well-known type-length-value scheme.
If both participants advertise the value, the extension is then enabled.
Usually, the extension defines new QUIC frames with dedicated type values.

To interact with transport parameters, we introduce two \protocoloperations: \pluginoperation{WriteTransportParameter} and \pluginoperation{DecodeTransportParameter}.
Each of these take as parameter the type of the transport parameter, i.e., an unsigned integer up to $2^{62}-1$.
Different plugins can then concurrently define new transport parameters having different type values.
To make the host implementation aware of these new types, plugins advertise during their initialization their \emph{transport parameter registrations}.
Such registrations enable \cquic to call \pluginoperation{WriteTransportParameter} with these new types.

 \begin{algorithm}[t]
 \caption{Protocol routines when sending frames.}
 \label{alg:sending-frames}
    \begin{algorithmic}
    \FORALL{frame reservations $r$}
        \IF{\pluginoperation{ShouldSendFrame}($r$\text{.frame\_type()})}
            \STATE $f \leftarrow$ \pluginoperation{PrepareFrame}($r$\text{.frame\_type()})
            \IF{\pluginoperation{FrameWireLen}($r$\text{.frame\_type()}), $f$) $\leq$ \text{buf\_len}}
                \IF{\pluginoperation{WriteFrame}($r$\text{.frame\_type()}), $f$) = \text{OK}}
                    \STATE \pluginoperation{OnFrameReserved}($r$\text{.frame\_type()}), $f$)
                \ENDIF
            \ENDIF
        \ENDIF
    \ENDFOR
    \end{algorithmic}
\end{algorithm}

Once negotiated, the plugin can then define frames with a new type.
Like transport parameters, the plugin advertises its \emph{frame registrations} to make \cquic aware of these types.
\cquic defines nine frame-related \protocoloperations, all taking as parameter the frame type.
Algorithm~\ref{alg:sending-frames} shows how these \protocoloperations operate when sending plugin-defined frames.
When preparing a packet to be sent, the \cquic implementation collects all the frame registrations made by the loaded plugins.
Then, for each registration, \pluginoperation{ShouldSendFrame} indicates whether the specified frame should be considered for sending in the current packet.
If so, \pluginoperation{PrepareFrame} internally prepares the frame data structure.
Note that this routine may also alter the sending processing, e.g., to stop
sending packets for some reason (see, e.g., Sect. \ref{sec:usecase-privacy-padding}).
Once the frame is prepared, \pluginoperation{FrameWireLen} returns its size on the wire to let the core
check that it can be written in a buffer of a given length.
Then, \pluginoperation{WriteFrame} %
writes the given frame in a buffer.
If the writing process succeeded, \cquic confirms through the \pluginoperation{OnFrameReserved} routine that the frame will be part of the sent packet.
Once the packet sent, the plugin can later be notified 
whether the frame has been either acknowledged or declared as lost thanks to the
\pluginoperation{NotifyFrame} routine.

The receiving process, on its side, mainly consists of two \protocoloperations.
The first, \pluginoperation{ParseFrame}, takes the raw bytes of the buffer and converts it into an internal processable structure.
The second, \pluginoperation{ProcessFrame}, uses the internal structure to adapt the state of the running connection.
Finally, the optional \pluginoperation{LogFrame} enables the plugin to provide a textual description of the frame when the core routine requires some logging.

One could note that, despite being generic, some \protocoloperations could have been merged.
For instance, \pluginoperation{PrepareFrame}, \pluginoperation{FrameWireLen}, and \pluginoperation{WriteFrame} could have been combined into a single \protocoloperation, as some QUIC implementations do. %
However, other implementations have different internals where the frame scheduling and the on-wire conversion are performed by distinct modules.
To ease its adoption, and because splitting a function into several smaller ones is usually a small effort, \cquic adopts this decomposed scheme.

In addition to the previous ones, \cquic defines three additional \protocoloperations.
\pluginoperation{Init} is used to initialize the plugin, e.g., to make registrations.
\pluginoperation{OnPluginTimeout} provides timer-based callbacks and are further described in Section~\ref{sec:running-plugins}.
\pluginoperation{PluginControl} serves as an external API, similar to \texttt{ioctl}, that can be called by other plugins or by the application. %

While extensions usually define new behaviors with new \protocoloperations, they may also want to monitor routines without directly altering them, e.g., for logging purposes.
Each \protocoloperation has three anchors.
The \anchor{define} anchor sets the behavior of the \protocoloperation, and at most one plugin can attach to this anchor.
The \anchor{before} and \anchor{after} anchors defines hooks when calling and returning from the \protocoloperation, respectively.
Multiple plugins can attach on the same \anchor{before} and \anchor{after} anchors.
However, unlike the \anchor{define} one, code loaded with these anchors cannot modify the \cquic host implementation state and does not output any value.

Finally, note that a plugin should not support a new frame if the extension is not supported by the peer.
To achieve this, \cquic splits the plugin loading into two steps.
Initially, only \pluginoperation{Init}, \pluginoperation{WriteTransportParameter} and \pluginoperation{DecodeTransportParameter} are loaded.
If the negotiation succeeds, or if no negotiation is needed, the plugin notifies \cquic that all the remaining routines can be loaded.

\subsection{Specifying a Common Representation}\label{sec:common-representation}
Since plugins extend or tune running connections, they need to access their state, such as whether the connection is a server-side one or what is the congestion window value.
To interact with \cquic implementations, plugins need a stable interface.
This %
not only covers the routines (as described in Sec.~\ref{sec:protocol-operations}), but also the exposed data structures.
For instance, from Alg.~\ref{alg:sending-frames}, \pluginoperation{PrepareFrame} outputs a frame structure that is then taken as input of \pluginoperation{FrameWireLen}, \pluginoperation{WriteFrame} and \pluginoperation{OnFrameReserved}.
This frame structure may have different internal representations on the \cquic host implementation side, but the plugin must have a stable layout.

To define such interface, \cquic introduces two elements in the \pluginopcommon library shared by both the host implementation and the plugin.
The first element consists in a list of all the QUIC fields that a plugin may access or modify.
Such fields derive from the core specification of QUIC~\cite{rfc9000} and relate to different elements such as connection-wide (server-side connection,...), packet number space (next number to send,...) and recovery fields (the specification~\cite{rfc9002} lists such relevant metrics).
As of February 2024, \cquic defines 43 fields.
Depending on the host implementation internals, some fields can be harder to collect. %
Instead of requiring a complete support, \cquic implementations can only provide access to a part of these fields.
When loading a plugin, \cquic checks that the host implementation supports all the requested fields.
If it is not the case, the plugin is denied loading.
Such an approach enables incremental \cquic compliance, hence easing its adoption.

The exposed fields contain different types.
For instance, whether the connection is server-side is associated to a boolean, whereas the next packet number to send corresponds to an unsigned integer.
Furthermore, \protocoloperations typically require inputs, as previously shown.
To address this, the second element of the \cquic interface consists in a list of data structure types that can be exchanged between the \cquic implementation and the plugin.
Besides primitive types (boolean, integer,...), it also provides time-related structures, socket address structures and QUIC-specific fields such as frames.
From the implementer perspective, the main effort is to write the conversion from the internal representation of the QUIC data structure to the \cquic one.
\cquic follows the mapping of the QUIC specification~\cite{rfc9000} for its plugin-exposed structure, so such effort is in practice limited.

Note that plugins may need to exchange raw bytes with the \cquic host implementation.
Because \cquic wants to keep isolation between the host implementation and the plugin, it introduces a specific \texttt{Bytes} type in the stable data structure list.
This type corresponds to a structure holding three fields: a tag referring to the raw bytes to exchange and the maximum length that can be read or written in these raw bytes, respectively.
Such an approach enables \cquic to provide a capability-based approach for interactions between the host implementation and the plugin.

\subsection{Running Plugins}\label{sec:running-plugins}

To dynamically extend the behavior of an implementation, \cquic introduces an environment in which the extension bytecode can be safely executed.
Such an environment must cope with two main concerns.
First, it should abstract the plugin from the actual computer system it runs.
Second, as plugins are bytecode external from the \cquic host implementation, it should provide isolation mechanisms as well as monitoring capabilities.

In the scope of software-based network protocol programmability, previous works~\cite{de_coninck_pluginizing_2019,wirtgen_xbgp_2023} relied on user-space version of the eBPF Virtual Machine (VM) due to its popularity in the Linux kernel~\cite{fleming_thorough_2017}.
While the in-kernel eBPF has now a mature support (verifier, BPF Type Format,...), it bases on features provided by the Linux kernel.
The integration for other applications requires consequent effort that is error-prone such as memory management and isolation.
Instead, to run plugins, \cquic relies on WebAssembly (\wasm)~\cite{haas_bringing_2017}.
Unlike eBPF, \wasm was designed with security in mind~\cite{dejaeghere2023comparing}.
While initially scoped for the web browsers, it is also empowering embedded devices~\cite{jacobsson2018virtual} and blockchain~\cite{eos} usecases.

A \cquic plugin consists in a \wasm module and contains several sections.
The most important ones are the code, the imported functions, the exported functions, and the memory.
The code section contains the instructions defining the plugin behavior.
The exported functions define entry-points where the module can be called.
The imported functions correspond to the required API for its execution. %
The module also has a specific linear memory %
storing its persistent data.

A plugin defines the different hooks it wants to attach thanks to its exported functions.
Specifically, the functions follow a naming convention to determine to which \protocoloperations they want to be attached.
When loading the plugin, \cquic's \pluginop checks the different plugin's exported functions and maps their name to the associated \protocoloperation.
For instance, a function with name \texttt{write\_frame\_aa} will be loaded at the \anchor{define} anchor of the \pluginoperation{WriteFrame} \protocoloperation with frame type 0xAA.
Similarly, the exported function \texttt{after\_process\_frame\_2} corresponds to the \anchor{after} anchor of the \pluginoperation{ProcessFrame} routine with type 2 (i.e., ACK frames).
Except their name, all exported functions follow the same signature by having a single integer input and a single integer output.
This output value notifies to the \cquic host implementation whether the plugin's function correctly performed its routine or failed with an error code.

\begin{table}[t]
    \caption{Summary of the \cquic API available to plugins.}
    \label{tab:cquic-api}
    \centering
    \begin{tabular}{|c|c|}
    \hline
      Category & Purpose \\
      \hline
      \hline
      Getters/Setters & Interact with \cquic session \\
      \hline
      {\scriptsize \texttt{get\_input/save\_output}} & Parameter communication \\
      \hline
      \texttt{Bytes} API & Raw bytes reading/writing \\
      \hline
      File system API & Read and write persistent files (logs,...) \\
      \hline
      Registration API & Notify new types to host implementation \\
      \hline
      \texttt{enable\_plugin} & Fully enable all routines (negotiated) \\
      \hline
      Time API & Provide timer-based callbacks \\
      \hline
    \end{tabular}
\end{table}

However, the plugin bytecode needs further support from \cquic to interact with the host implementation.
All these required features are listed in the imported functions of the plugin.
To ease their development,
\cquic introduces to plugins the \pluginopwasm library that takes care of listing the required import functions.
Furthermore, it offers a safe interface to plugins by abstracting the Foreign Function Interface (FFI) internals and makes \cquic data structures available through the \wasm linear memory.
The offered plugin's API is summarized in Table~\ref{tab:cquic-api}.
It includes getting and setting \cquic session's state, fetching and saving protocol routine-specific inputs and outputs, and reading/writing raw bytes or files.
The \cquic API also provides functions to register transport parameter and frame types and to fully activate the plugin (see Sec.~\ref{sec:protocol-operations}).
Finally, the \cquic API enables plugin to register timers along with callback functions.
These timers are then considered in the \cquic host implementation to let a plugin wake the stack.

However, given the external nature of the plugin bytecode, \cquic implementations may want to restrict parts of the API made available.
For instance, a host implementation may prevent untrusted plugins from accessing sensitive session's fields, such as cryptographic keys.
\cquic adopts a capability-based permission system whose rights are determined by the host implementation.
Different permissions can be provided on a per-plugin basis, enabling different levels of trust in the loaded plugins.
If a plugin tries to access a forbidden field or API, the running environment will deny access by returning an error code to the plugin.

\subsection{Making Implementations \cquic-Compliant}\label{sec:making-cquic-compliant}

To support \cquic, a QUIC host implementation needs to integrate the three previously discussed elements: $i)$ turning internal functions into \protocoloperations, $ii)$ adding conversion functions from internal data structures to \cquic ones, and $iii)$ embedding the plugin environment runtime.
The \pluginop library eases this process by fully taking care of $iii)$, remaining for the implementer to focus on $i)$ and $ii)$.

To turn an internal function (e.g., the one processing incoming frames) into a \protocoloperation, the implementer performs the following steps.
First, she wraps the content of the exposed function in an internal function.
Then, the original function first checks if a plugin is attached to the related \protocoloperation with the given parameter.
If so, the parameters of the functions are converted to the \cquic representation to be processable by the plugins.
Related plugin code can then be called at their related anchors thanks to the \pluginop library.
If the \protocoloperation outputs values, the \cquic data structures are then converted back to the host implementation ones.
Note that if there is no plugin for the \protocoloperation with the \anchor{define} anchor, the function then calls the wrapped internal function holding the original function behavior.

While the code performing data structure conversions is typically a purely additive change, the remaining changes are not.
To limit the codebase changes, and indirectly human mistakes, \cquic comes with a \pluginopmacro library defining macros that automate such code changes at compilation time.
Turning a function into a \protocoloperation is then made with a single line of code, making \cquic adoption easier.
Note that the implementer may need to add new code, e.g., to support sending new frames as depicted by Algorithm~\ref{alg:sending-frames}, but without requiring modification of the remaining code base.

\section{Implementation}\label{sec:implementation}

\begin{figure}[t]
	\centering
	\includegraphics[width=\columnwidth]{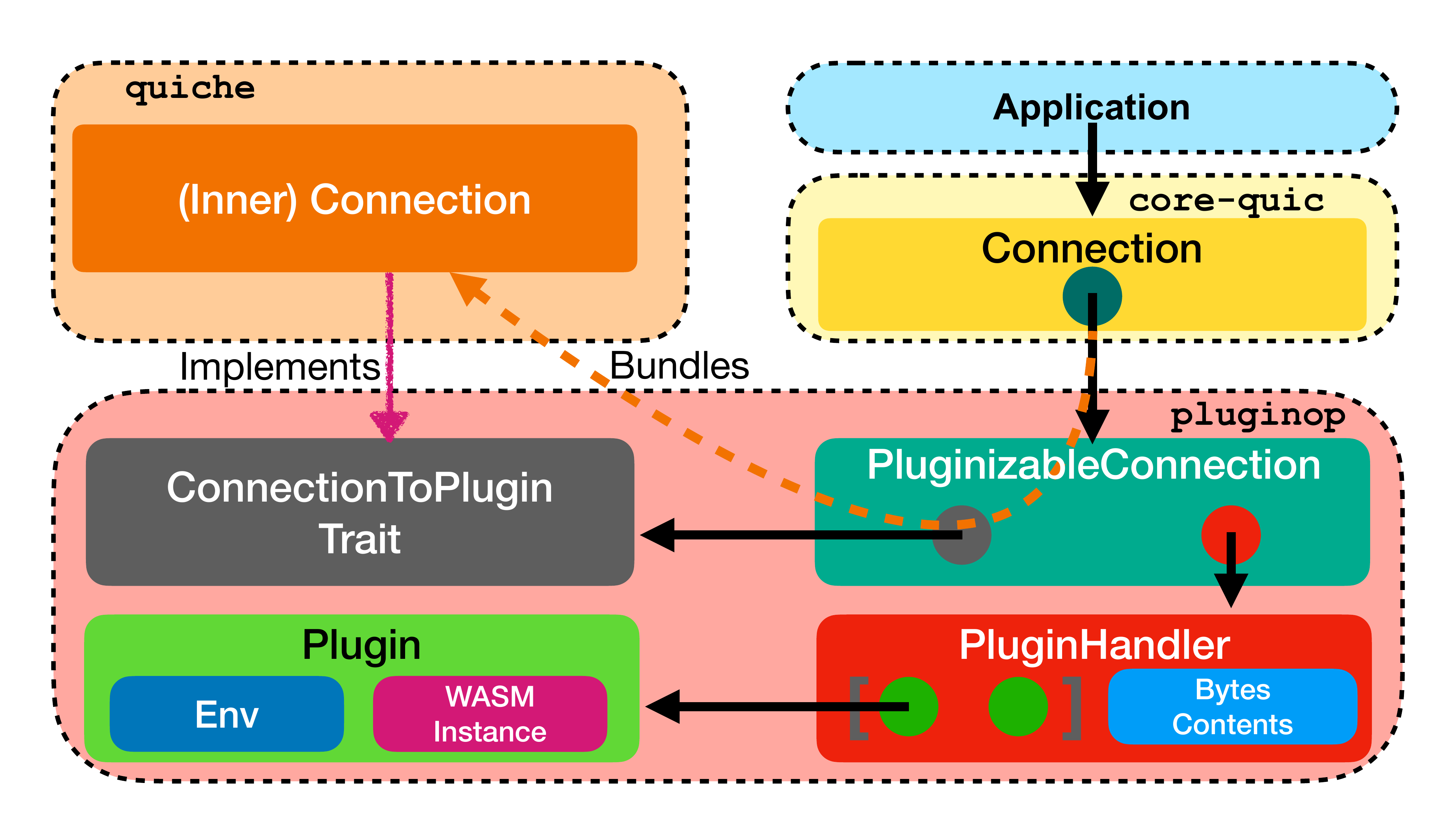}
	\caption{The architecture of \pluginop. Black arrows mean "contains".}
	\label{fig:engine-arch}
\end{figure}

To make \cquic concrete, we implemented the different architectural components depicted in Figure~\ref{fig:overview-architecture}.
We rely on the Rust language~\cite{matsakis_rust_2014} as it enforces safety in performance-critical software. %
Each of the blocks in Figure~\ref{fig:overview-architecture} has been implemented as a Rust library.
To make an implementation \cquic-compliant, it needs to integrate the \pluginop library in its code base.
Figure~\ref{fig:engine-arch} depicts the internal architecture of the \pluginop library.
To make a session pluginizable, the implementer needs to wrap the connection structure into a \pluginop's \texttt{PluginizableConnection}.
This structure contains two fields.
The first is a \texttt{PluginHandler} that handles all the processing required by the dynamic pluginization.
The second is the original connection structure itself.
Note that the \texttt{PluginizableConnection} requires the original connection structure to implement the \texttt{ConnectionToPlugin} trait. %
This trait contains two methods that establishes a common interaction between the pluginization engine and the QUIC implementation.
Concretely, these allow getting and setting connection fields (as defined by \pluginopcommon) based on a serialized value.
Note that this original connection structure is agnostic to whether plugins are loaded or not.

The \texttt{PluginHandler} structure takes care of the pluginization process.
It is generic with relation to the actual QUIC implementation, as it only operates on the \texttt{ConnectionToPlugin} trait.
When a plugin wants to augment a \cquic session, it is inserted in the \texttt{PluginHandler}.
A \texttt{Plugin} structure is created and contains the \wasm instance along with a plugin-specific environment (e.g., the permission it has regarding fields and API it can access, its inputs/outputs,...).
Our \pluginop library relies on the \texttt{wasmer} \wasm engine~\cite{wasmer} to execute the plugin bytecode, and on its Singlepass \wasm compiler.
When calling a \protocoloperation, \texttt{PluginHandler} checks if a \texttt{Plugin} provides the requested routine in \anchor{define} anchor.
If so, the related bytecode is then executed, and its result given back to the caller.
Furthermore, the \pluginop library provides the API required by the plugins to be run.

The common structures are provided by the \pluginopcommon library.
It notably defines the only type, \texttt{PluginVal}, communicated between the host implementation and plugins.
Practically, \texttt{PluginVal} is an \texttt{enum} whose each variant corresponds to a concrete, operable type.
Similarly, the list of connection fields that a plugin can interact with is also implemented as an \texttt{enum}.

While plugins correspond to \wasm modules, their source code may be in any language.
However, relying on unsafe languages such as C may lead to exploitable \wasm bytecodes~\cite{lehmann2020everything}.
A whole range of safety efforts can be prevented, by design, by relying on Rust.
Therefore, we propose to plugin developers to write their source code in safe Rust. %
Yet, the \wasm external functions rely on the Foreign Function Interface (FFI) which is, by essence, unsafe to use.
To mitigate this, the \pluginopwasm Rust library handles all the serialization concerns related to the FFI and provides a safe abstraction on which plugins can be built.
In addition, the \pluginopwasm exposes the \cquic data structures defined by the \pluginopcommon library.

\begin{table}[t]
    \caption{Summary of Code Contribution, Excluding Comments and Whitespaces.}
    \label{tab:summary_loc}
    \centering
    \begin{tabular}{|c|c|}
    \hline
        Component & \# Lines of Rust code \\
    \hline
    \hline    
        Core host-side library \pluginop & 1700 \\
    \hline
        Macro helpers \pluginopmacro & 400 \\
    \hline
        WebAssembly-side library \pluginopwasm & 430 \\
    \hline
        Common structure library \pluginopcommon & 730 \\
    \hline
    \end{tabular}
\end{table}

Table~\ref{tab:summary_loc} quantifies the implementation effort of \cquic. 
Overall, the generic libraries represent about 3250 lines of Rust code.
Most of the complexity lies in the \pluginop library.
The \pluginopcommon library mostly defines structures and enumerations to be used by other libraries and the plugins.

\subsection{Integration into QUIC Implementations}

To demonstrate the generality of \cquic's \protocoloperations and the applicability of \pluginop, we make two QUIC implementation \cquic-compliant.
First, we focus on the Cloudflare's \texttt{quiche} implementation providing a library to be integrated in networking applications to perform QUIC.
Second, we look at the \texttt{quinn} implementation offering a complete QUIC network abstraction for applications.
Due to the QUIC complexity, many implementations abstract the I/O interactions from the QUIC logic.
This is the case for the \texttt{quiche} implementation, and \texttt{quinn} delegates the protocol logic handling in the \texttt{quinn-proto} library.
Similar approaches are also taken by other QUIC implementations, such as Amazon's \texttt{s2n-quic} and Mozilla's \texttt{neqo}.

Ideally, the integration effort of \cquic into existing implementations should be low.
This translates into two concrete properties.
First, the public API offered by these stacks should not introduce any breaking changes, such as structure or function renaming.
Second, the code diff should be as low as possible, and with only purely additive changes when possible.
Figure~\ref{fig:engine-arch} shows how \pluginop integrates with the \texttt{quiche} implementation.
To make it pluginizable, its \texttt{Connection} structure needs to implement the \texttt{ConnectionToPlugin} trait and some of its internal functions should be converted to \protocoloperations thanks to macros provided by the \pluginopmacro library.
Overall, this introduces a code diff of +1043, -46 over a code base of 38000 lines of Rust code.
More than half of the additive changes are related to conversion code between internal data structures and \cquic's ones.
To let applications benefit from the \cquic-compliant \texttt{quiche} version, we wrote a 100-line \texttt{core-quic} library exposing public \texttt{quiche} API along with functions to insert plugins on a per-session basis.
These applications then simply need to rely on the \texttt{core-quic} library instead of the \texttt{quiche} without further code modification.
Such results suggest reasonable efforts to deploy \cquic in \texttt{quiche}.

Making the \texttt{quinn} implementation \cquic-compliant only requires changes in the internal \texttt{quinn-proto} library, leading to a code diff of +1184, -253.
These are still mostly additive changes, and the large negative change is due to the wrapping of 200 lines of code into a dedicated function.
Yet, such changes remain reasonable for \cquic integration.

\section{Micro-benchmarks}\label{sec:microbenchmarks}

\cquic provides increased protocol flexibility and tuning.
Still, the pluginization mechanisms introduced by \cquic, i.e., providing new behaviors by the dynamic injection of \wasm bytecode, brings some computational overhead.
To quantify it, we focus on the \pluginop library and perform various benchmarks to quantify the overhead brought by each of provided features.
These benchmarks were performed using the Rust's \texttt{criterion} benchmark library.
Table~\ref{tab:micro-benchmarks} reports the median time over 100 runs of each benchmark after a warm-up phase of 3 seconds running on an Apple MacBook Pro M1 with 16 GB of RAM.

\begin{table}[t]
    \caption{\pluginop's micro-benchmarks.}
    \label{tab:micro-benchmarks}
    \centering
    \begin{tabular}{|c|c|}
        \hline
        Benchmark name & Median time \\
        \hline
        \hline
        \textbf{A} - Empty \texttt{wasmer} call & \SI{37.5}{\nano\second} \\
        \hline
        \textbf{B} - Empty plugin call & \SI{75}{\nano\second} \\
        \hline
        \textbf{C} - Plugin arithmetic operations & \SI{204}{\nano\second} \\
        \hline
        \textbf{D} - Set inputs and get outputs & \SI{710}{\nano\second} \\
        \hline
        \textbf{E} - Native send and receive & \SI{171}{\nano\second} \\
        \hline
        \textbf{F} - Plugin send and receive & \SI{2.46}{\micro\second} \\
        \hline
        \textbf{G} - Full MAX\_DATA plugin loading & \SI{1.36}{\milli\second} \\
        \hline
    \end{tabular}
\end{table}

Our first benchmark (\textbf{A}) calls through \texttt{wasmer} a \wasm function that directly returns.
This base overhead is due to the trampoline processing when switching from the native execution environment to the \wasm one.
Our second benchmark (\textbf{B}) consists in calling the same function through \pluginop.
Compared to \textbf{A}, there is additional overhead due to fetching the \wasm function from the \texttt{Plugin} structure and clearing previous \texttt{PluginVal} inputs and outputs.
Our third experience (\textbf{C}) calls two plugin functions.
The first call sets two integers directly in the plugin's memory.
Then, the second plugin call computes the four basic arithmetic operations on these stored integers and provides the result back to the caller.
Compared to \textbf{B}, the processing time increase comes from the call to two functions and the processing overhead of the computations. %
Our fourth experience (\textbf{D}) performs the same operations as \textbf{C}, except that inputs and outputs are now passed using the \texttt{PluginVal} common type.
The increased overhead is caused by the serialization and the deserialization of the two inputs and the four outputs between the host implementation and the \wasm environment (about \SI{80}{\nano\second} per parameter).
These operations also involve memory copies of the \texttt{PluginVal} structure into the plugin's linear memory that are not present in the previous benchmark.

With the understanding of these base tests, we now extend our benchmarks to the more representative use case of sending and receiving frames.
We implement a mocked \texttt{Connection} structure that simply sends frames following Algorithm~\ref{alg:sending-frames} and processes the received frames as described in Section~\ref{sec:protocol-operations}.
The whole process involves seven \protocoloperations. %
The first benchmark variant (\textbf{E}) consists in the native execution of both the sending and the receiving process when only a MAX\_DATA frame is required.
The second one (\textbf{F}) corresponds to the same scenario, except that all the required \protocoloperations are provided by a plugin.
This, considering the parameter communication, %
suggests a mean per-plugin call overhead of \SI{350}{\nano\second}, which is in line with the previous results.

Finally, there is overhead when loading plugins, i.e., the compilation and the instantiating of a \wasm bytecode by the execution engine.
When relying on the Singlepass compiler, the plugin loading is in the order of the millisecond (\textbf{G}).
This overhead only happens at the beginning of the session and caching techniques may be applied to reduce it.
Note that we could have relied on a compiler providing more efficient compiled code.
While the LLVM-based compiler enables \protocoloperation calls to be about 25~\% faster according to our benchmarks, the plugin loading time faces a 50$\times$ increase.

\section{Exploring Use Cases}\label{sec:usecase}

\begin{table}[t]
    \caption{Statistics about the implemented plugins.}
    \label{tab:usecases-loc}
    \centering
    \begin{tabular}{|c|c|c|}
    \hline
    Use case & \# Rust LoC & Bytecode size \\
    \hline
    \hline
      ACK logger (\S~\ref{sec:usecase-ack-logging}) & 74 & 35.9 KB \\
    \hline
      Privacy padding (\S~\ref{sec:usecase-privacy-padding}, \S~\ref{sec:usecase-combined}) & 131 & 43.3 KB \\
    \hline
      Probe path (\S~\ref{sec:usecase-probe-path}) & 112 & 48.9 KB \\
    \hline
      BDP Frame (\S~\ref{sec:usecase-bdp-frame}) & 170 & 55.8 KB \\
    \hline
      DUMMY frame (\S~\ref{sec:usecase-combined}) & 192 & 63.0 KB \\
    \hline
      
    \end{tabular}
\end{table}

This section now provides some examples of use cases that \cquic can currently provide.
Each of the discussed use case has been tested on our two \cquic-compliant implementations, namely \texttt{quiche} and \texttt{quinn}.
Table~\ref{tab:usecases-loc} provides some statistics about the implemented plugins that are discussed in this Section.
Note that each of the plugin bytecode have a size of at least 30 KB.
This is due to the (de)serialization code required to process the \texttt{PluginVal} structure.
Further optimization is possible but is left as future work.
All these plugins work with both our \cquic-compliant implementations based on \texttt{quiche} and \texttt{quinn}.
Unless otherwise stated, we rely on the example client and server provided by each implementation.

\subsection{Frame Logging}\label{sec:usecase-ack-logging}

Because of its encrypted nature, it is very hard to analyze a QUIC network trace without having access to the encryption keys.
Usually, QUIC implementations can output logs~\cite{marx2020debugging} to let operators debug their sessions.
\cquic can also help in this process by embedding the logging process in a plugin.
Such a plugin relies on the \anchor{before} and \anchor{after} anchors of \protocoloperations to log the different events in a file.
We implemented a prototype of such a logger reporting ACK frame events, showing that \cquic can provide monitoring capabilities without altering the core QUIC behavior.

\subsection{QUIC-based Privacy Solution}\label{sec:usecase-privacy-padding}

Although QUIC encrypts all the control and application data, it is still exposed to privacy concerns.
Packet sizes and timestamps are side-channel information that can be leveraged to build, e.g., website fingerprinting tools~\cite{rahman2020tik,smith2022qcsd}.
The observer can map a website by translating the network trace into a feature vector.
A classification algorithm then determines if the feature vector is similar to a website's one.

To counter such fingerprinting, endpoints could adopt website-fingerprinting defenses consisting in padding and delaying packets sent by the participating entities.
To show how \cquic can contribute to this field, we built a plugin that pads all packets to the MTU size (using PADDING frames) and adds random delay between sent packets.
To achieve such packet delaying, \cquic relies on the plugin return value of the \pluginoperation{PrepareFrame} routine to halt the batch of packet sending.
The exact delaying is then controlled by the Time API provided to the plugin as suggested by Table~\ref{tab:cquic-api}.
When the plugin timer fires, \cquic invokes the \pluginoperation{OnPluginTimeout} behavior provided by the plugin, which then enables further packet sending.
Note that this plugin, while altering the sending logic, does not introduce any extension frame.
Therefore, it can be enabled without handshake negotiation and such a feature can be deployed on only one participating endpoint.

\begin{figure}
    \centering
    \includegraphics[width=\columnwidth]{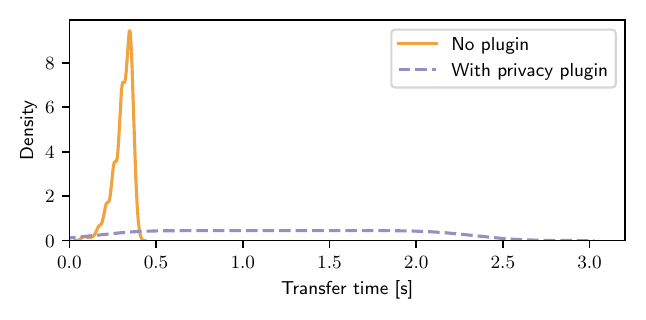}
    \caption{Time distribution of client's packets of a 500 KB download.}
    \label{fig:privacy-client-ts}
\end{figure}

\begin{figure}
    \centering
    \includegraphics[width=\columnwidth]{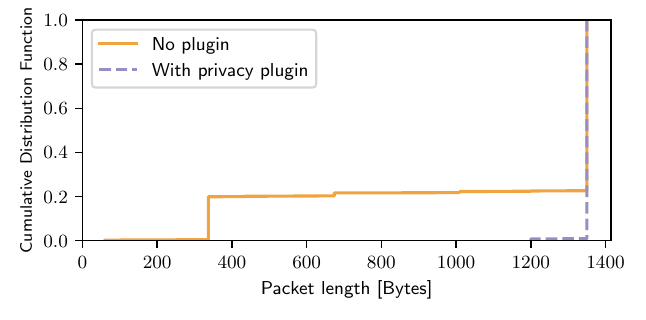}
    \caption{Size distribution of server's packets of a 500 KB download.}
    \label{fig:privacy-server-size}
\end{figure}

To evaluate its impact, we perform a Mininet~\cite{handigol_reproducible_2012} evaluation between a \texttt{quinn} client and a \texttt{quiche} server connected through a 50 Mbps, \SI{40}{\milli\second} RTT network path.
We study a file download of 500 KB.
We consider two setups.
First, the regular behavior of the unmodified implementations.
Second, the situation where both the \texttt{quinn} client and the \texttt{quiche} server are tuned by the same \cquic plugin that pads and delays sent packets.
For each variant, we perform 30 runs.
Figure~\ref{fig:privacy-client-ts} shows that the unmodified time distribution of the client's sent packets is driven by the received server's packets whose follow the typical Cubic's slow start increase.
The random delay introduced by the plugin breaks this pattern by flattening and expanding the probability density function curve.
A similar result is observed for the time distribution of server's sent packets.
Besides, Figure~\ref{fig:privacy-server-size} shows that server's packets have their size following the congestion control limitation.
The injected plugin hides such information by padding all the packets\footnote{Except the QUIC Initial packets being 1200-byte long.} to the MTU size of the network path, i.e., 1350 bytes.
This shows the potential that \cquic brings to develop dynamic countermeasures to such traffic fingerprinting.

\subsection{Application-controlled Network Probing}\label{sec:usecase-probe-path}

Existing QUIC implementations provide different APIs.
An application may want a feature that its serving implementation does not give access.
As an example, an application facing an idling connection may need to probe its network path.
There is two ways to achieve this in QUIC.
A first way consists in sending a PING frame, which elicits an ACK frame from the peer.
The second one implies sending a PATH\_CHALLENGE frame and the peer replies with a PATH\_RESPONSE one.
Our prototype plugin implements the second approach.
It provides a \pluginoperation{PluginControl} API where the application can request the sending of the probe.
Another \pluginoperation{PluginControl} entry point enables the application to get the experienced latency between the PATH\_CHALLENGE and the PATH\_RESPONSE.

\subsection{Optimizing QUIC in Large RTT Scenarios}\label{sec:usecase-bdp-frame}

QUIC brings interests to operate it in challenging environment such as satellite or interplanetary communications.
While such environments may provide large bandwidth, the distance between endpoints implies a consequent latency affecting the communication's experience.
The congestion control's slow-start process usually takes several RTTs before operating on the available bandwidth, implying large delays for short transfers.
A recent proposal~\cite{kuhn-quic-bdpframe-extension-05} suggests defining a frame where the server advertises to the client its congestion control state to properly operate in such environment.
When initiating a new connection to the same server, the client can then send back the frame to the server to allow it to resume its congestion control state and directly operate at the target rate.

\begin{figure}
    \centering
    \includegraphics[width=\columnwidth]{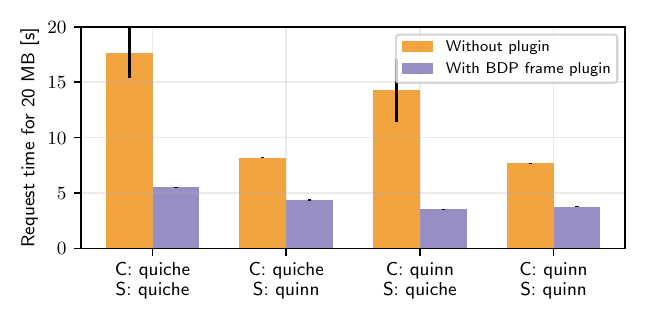}
    \caption{The BDP frame plugin enables the server to directly operate to a suitable operating state.}
    \label{fig:bdp-frame}
\end{figure}

We implement the proposed BDP frame idea in a \cquic plugin.
Because it introduces new mechanisms, it requires negotiation using the QUIC transport parameters, and the plugin is only enabled when the peer advertises its support.
This plugin covers all transport parameters and frames \protocoloperations.
To evaluate its potential benefits, we consider a Mininet scenario where endpoints are connected to a 50 Mbps, \SI{500}{\milli\second} RTT network path.
We consider the four possible setups where the \texttt{quinn} and \texttt{quiche} implementations operate as the client and/or the server.
For each case, we compare the default behavior of the implementations with the one provided by our plugin.
Each distinct case is run 20 times.
Figure~\ref{fig:bdp-frame} shows the median and the standard deviation of the transfer time to download a 20 MB file.
In such a large RTT scenario, a lot of time is wasted in probing the network capacity and the transfer completes when the \texttt{quinn} server finally reaches its target rate.
Our experiments report that the Cubic implementation in the \texttt{quiche} server badly behaves in such cases and often exits the slow start phase without reaching the optimal value.
When the BDP frame is introduced by the plugin, the server can directly reuse the optimal congestion control state for the scenario and the transfer time considerably decreases in all implementation configurations.

\subsection{Combining Extensions}\label{sec:usecase-combined}

One of the \cquic design goals is to support the inclusion of multiple plugins providing orthogonal extensions.
All the previously discussed plugins can be injected over a single connection to merge their functionalities.
But plugins can also cooperate between each other's.
To illustrate this, we setup the following toy case. 
The privacy padding plugin has a \pluginoperation{PluginControl} routine letting the caller to dynamically enable the packet padding and delaying features over the session.
We then define a new plugin introducing a DUMMY frame.
If after the handshake both endpoints support it, the plugin makes the host send this new frame, containing no data, only when there is no other DUMMY frame in-flight.
This implements a once-per-perceived-RTT sending logic.
Every four received DUMMY frame, the endpoint calls the \pluginoperation{PluginControl} routine related to the privacy padding plugin to toggle its activation.
Note that if the routine is not provided, e.g., because the privacy padding plugin is not loaded over the connection, the caller is simply notified that the behavior is not available.

\begin{figure}
    \centering
    \includegraphics[width=\columnwidth]{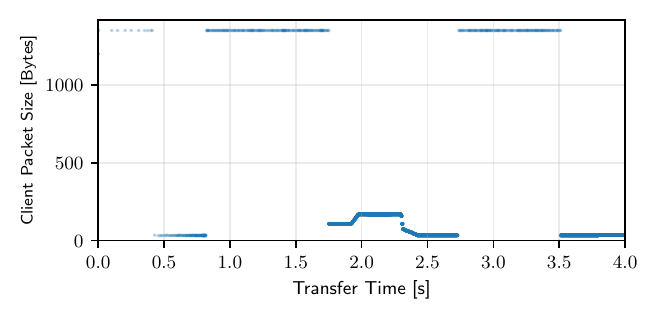}
    \caption{Occurrences of client's packets when combining the DUMMY frame extension with the privacy padding one.}
    \label{fig:combined}
\end{figure}

To visualize this feature, we consider a scenario where the \texttt{quiche} client loads both the privacy padding and the DUMMY frame plugins, while the \texttt{quinn} server only inserts the DUMMY frame one.
We consider a scenario where endpoints are connected by a 50 Mbps, \SI{100}{\milli\second} RTT network path.
Figure~\ref{fig:combined} shows the time and the size of packets sent by the client for a given run.
Initially, the privacy plugin pads the client's packets to the MTU size.
After about \SI{400}{\milli\second}, the client received four DUMMY frames from the server and disabled the privacy plugin, leading to much smaller packets.
Then, about \SI{400}{\milli\second} later, the privacy plugin is enabled again.
This toggling continues until the end of the transfer, but this later tends to happen every second.
This is due to the Cubic's server congestion control that saturates the path, and the network buffer introduces queuing delay, increasing %
the perceived RTT.
The client packets between \SI{1.75}{\second} and \SI{2.5}{\second} contains large ACK frames with up to 70 ACK blocks.
These illustrate packet losses due to the saturation of the network buffer.
Our results thus demonstrate that such plugins can be isolated while providing a cooperative interface between them.

\section{Related Works}\label{sec:related}
In the late 90s, there were a strong interest in building active networks where network protocol operations could be extended by letting network nodes execute bytecode included in the exchanged packets~\cite{tennenhouse_survey_1997,hicks_plan:_1998}.
However, such approaches faced limited deployment, notably due to security considerations~\cite{psounis_active_1999}.
Still, this idea of reconfigurable network nodes led to the development of more mature solutions such as Software Defined Networks~\cite{mckeown_openflow:_2008} and P4~\cite{bosshart_p4:_2014}.
These approaches target network layer solutions and usually require specific hardware.
At the transport layer, there were previous attempts in making configurable and extensible transport protocols~\cite{bridges_configurable_2007,patel_upgrading_2003}.
These were not widely deployed for several reasons (unencrypted protocols, very specific implementation architecture,...).
With the rise of the QUIC protocol, researchers~\cite{de_coninck_pluginizing_2019} proposed PQUIC to dynamically extend a QUIC implementation with plugins executed by a virtual machine.
However, plugins are tight to the researchers' implementation and PQUIC relies on a core hash-table holding all exposed operations, requiring large implementation changes.
More recently, Wirtgen et al.~\cite{wirtgen_xbgp_2023} proposed xBGP, a solution to dynamically extend BGP implementations with a same plugin.
While xBGP has similar ideas, \cquic focuses on the QUIC protocol and relies on different building blocks, e.g., to run plugins.
There are also orthogonal works applying dynamic programmability in other encrypted protocols~\cite{rochet_tcpls_2021,rochet2022towards}.

\section{Conclusion}\label{sec:conclusion}

This paper presented \cquic, an extension to the QUIC standard that enables its compliant implementations to be dynamically extended by a common, implementation-agnostic plugin.
Once deployed, \cquic nodes provides protocol tuning and extensibility without requiring any change to the implementation's binary.
We demonstrated that such an approach is deployable by making two different QUIC implementations \cquic-compliant with limited code base changes.
We also showcased a few use cases where a same plugin can extend both \cquic-compliant implementations.

We believe \cquic addresses the transport protocol extensibity inertia.
While it can be deployed at both sides, \cquic is well-suited for client implementations, often being out of control, and server can then natively implement extensions they want to support.
Our future works include making more QUIC implementations \cquic-compliant to strengthen the genericity of the API and extending \cquic to cover more complex use cases requiring more than operations on frames.

\textbf{Artifacts available:} \url{https://core-quic.github.io}.

\bibliographystyle{IEEEtran}
\bibliography{IEEEabrv,bib}

\begin{thebibliography}{10}
\providecommand{\url}[1]{#1}
\csname url@samestyle\endcsname
\providecommand{\newblock}{\relax}
\providecommand{\bibinfo}[2]{#2}
\providecommand{\BIBentrySTDinterwordspacing}{\spaceskip=0pt\relax}
\providecommand{\BIBentryALTinterwordstretchfactor}{4}
\providecommand{\BIBentryALTinterwordspacing}{\spaceskip=\fontdimen2\font plus
\BIBentryALTinterwordstretchfactor\fontdimen3\font minus
  \fontdimen4\font\relax}
\providecommand{\BIBforeignlanguage}[2]{{%
\expandafter\ifx\csname l@#1\endcsname\relax
\typeout{** WARNING: IEEEtran.bst: No hyphenation pattern has been}%
\typeout{** loaded for the language `#1'. Using the pattern for}%
\typeout{** the default language instead.}%
\else
\language=\csname l@#1\endcsname
\fi
#2}}
\providecommand{\BIBdecl}{\relax}
\BIBdecl

\bibitem{honda_is_2011}
M.~Honda \emph{et~al.}, ``\BIBforeignlanguage{en}{Is {It} {Still} {Possible} to
  {Extend} {TCP}?}'' in \emph{\BIBforeignlanguage{en}{{ACM} {IMC} '11}}, 2011,
  pp. 181--194.

\bibitem{budzisz_taxonomy_2012}
L.~Budzisz \emph{et~al.}, ``A taxonomy and survey of {SCTP} research,''
  \emph{ACM Computing Surveys (CSUR)}, vol.~44, no.~4, p.~18, 2012.

\bibitem{wirtgen_xbgp_2023}
T.~Wirtgen \emph{et~al.}, ``{xBGP}: Faster innovation in routing protocols,''
  in \emph{USENIX NSDI'23}, Apr. 2023, pp. 575--592.

\bibitem{rfc9000}
J.~Iyengar and M.~Thomson, ``{QUIC: A UDP-Based Multiplexed and Secure
  Transport},'' RFC 9000, May 2021.

\bibitem{rfc8446}
E.~Rescorla, ``{The Transport Layer Security (TLS) Protocol Version 1.3},'' RFC
  8446, Aug. 2018.

\bibitem{de_coninck_pluginizing_2019}
Q.~De~Coninck \emph{et~al.}, ``\BIBforeignlanguage{en}{Pluginizing {QUIC}},''
  in \emph{\BIBforeignlanguage{en}{{ACM} {SIGCOMM} '19}}.\hskip 1em plus 0.5em
  minus 0.4em\relax Beijing, China: ACM Press, 2019, pp. 59--74.

\bibitem{rfc9002}
J.~Iyengar and I.~Swett, ``{QUIC Loss Detection and Congestion Control},'' RFC
  9002, May 2021.

\bibitem{fleming_thorough_2017}
M.~Fleming, ``A thorough introduction to {eBPF},'' \emph{Linux Weekly News},
  Dec. 2017, \url{https://old.lwn.net/Articles/740157/, Accessed: 2021-02-04},.

\bibitem{haas_bringing_2017}
A.~Haas \emph{et~al.}, ``Bringing the web up to speed with {WebAssembly},''
  \emph{ACM SIGPLAN Notices}, vol.~52, no.~6, pp. 185--200, 2017.

\bibitem{dejaeghere2023comparing}
J.~Dejaeghere \emph{et~al.}, ``Comparing security in ebpf and webassembly,'' in
  \emph{ACM Workshop on eBPF and Kernel Extensions}, 2023, pp. 35--41.

\bibitem{jacobsson2018virtual}
M.~Jacobsson and J.~Will{\'e}n, ``Virtual machine execution for wearables based
  on webassembly,'' in \emph{EAI BODYNETS}, 2018, pp. 381--389.

\bibitem{eos}
\BIBentryALTinterwordspacing
eosio, ``Eos virtual machine: A high-performance blockchain webassembly
  interpreter,'' 2019. [Online]. Available:
  \url{https://eos.io/news/eos-virtual-machine-a-high-performance-blockchain-webassembly-interpreter/}
\BIBentrySTDinterwordspacing

\bibitem{matsakis_rust_2014}
N.~D. Matsakis and F.~S. Klock~II, ``The rust language,'' \emph{ACM SIGAda Ada
  Letters}, vol.~34, no.~3, pp. 103--104, 2014.

\bibitem{wasmer}
\BIBentryALTinterwordspacing
WasmerIO. Wasmer. [Online]. Available: \url{https://github.com/wasmerio/wasmer}
\BIBentrySTDinterwordspacing

\bibitem{lehmann2020everything}
D.~Lehmann \emph{et~al.}, ``Everything old is new again: Binary security of
  {WebAssembly},'' in \emph{USENIX Security '20}, 2020, pp. 217--234.

\bibitem{marx2020debugging}
R.~Marx \emph{et~al.}, ``Debugging quic and http/3 with qlog and qvis,'' in
  \emph{ACM ANRW '20}, 2020, pp. 58--66.

\bibitem{rahman2020tik}
M.~S. Rahman \emph{et~al.}, ``Tik-tok: The utility of packet timing in website
  fingerprinting attacks,'' \emph{PoPETs '20}, vol.~3, pp. 5--24, 2020.

\bibitem{smith2022qcsd}
J.-P. Smith \emph{et~al.}, ``{QCSD}: A {QUIC} client-side
  website-fingerprinting defence framework,'' in \emph{USENIX Security '22},
  pp. 771--789.

\bibitem{handigol_reproducible_2012}
N.~Handigol \emph{et~al.}, ``\BIBforeignlanguage{en}{Reproducible {Network}
  {Experiments} {Using} {Container}-{Based} {Emulation}},'' in
  \emph{\BIBforeignlanguage{en}{ACM CoNEXT '12'}}, pp. 253--264.

\bibitem{kuhn-quic-bdpframe-extension-05}
N.~Kuhn \emph{et~al.}, ``{Signalling CC Parameters for Careful Resume using
  QUIC},'' IETF Draft draft-kuhn-quic-bdpframe-extension-05, Mar. 2024.

\bibitem{tennenhouse_survey_1997}
D.~L. Tennenhouse \emph{et~al.}, ``A survey of active network research,''
  \emph{IEEE communications Magazine}, vol.~35, no.~1, pp. 80--86, 1997.

\bibitem{hicks_plan:_1998}
M.~Hicks \emph{et~al.}, ``{PLAN}: {A} packet language for active networks,''
  \emph{ACM SIGPLAN Notices}, vol.~34, no.~1, pp. 86--93, 1998.

\bibitem{psounis_active_1999}
K.~Psounis, ``Active networks: {Applications}, security, safety, and
  architectures,'' \emph{IEEE Communications Surveys}, vol.~2, no.~1, pp.
  2--16, 1999.

\bibitem{mckeown_openflow:_2008}
N.~McKeown \emph{et~al.}, ``{OpenFlow}: enabling innovation in campus
  networks,'' \emph{ACM SIGCOMM CCR}, vol.~38, no.~2, pp. 69--74, 2008.

\bibitem{bosshart_p4:_2014}
P.~Bosshart \emph{et~al.}, ``P4: {Programming} protocol-independent packet
  processors,'' \emph{ACM SIGCOMM CCR}, vol.~44, no.~3, pp. 87--95, 2014.

\bibitem{bridges_configurable_2007}
P.~G. Bridges \emph{et~al.}, ``A {Configurable} and {Extensible} {Transport}
  {Protocol},'' \emph{IEEE/ACM ToN}, vol.~15, no.~6, pp. 1254--1265, Dec. 2007.

\bibitem{patel_upgrading_2003}
P.~Patel \emph{et~al.}, ``Upgrading {Transport} {Protocols} using {Untrusted}
  {Mobile} {Code},'' \emph{ACM SIGOPS OSR}, vol.~37, no.~5, pp. 1--14, 2003.

\bibitem{rochet_tcpls_2021}
F.~Rochet \emph{et~al.}, ``Tcpls: Modern transport services with tcp and tls,''
  in \emph{ACM CoNEXT'21}, 2021, p. 45–59.

\bibitem{rochet2022towards}
F.~Rochet and T.~Elahi, ``Towards flexible anonymous networks,'' \emph{arXiv
  preprint arXiv:2203.03764}, 2022.

\end{thebibliography}

\end{document}